\documentclass{sig-alternate.fix-ref-space}
\usepackage{color}
\usepackage{graphicx}
\usepackage{xspace}
\title{Respect My Authority!  HITS Without Hyperlinks, Utilizing Cluster-Based Language Models}
\numberofauthors{1}
\author{\alignauthor Oren Kurland and Lillian Lee\\
\affaddr{Department of Computer Science}\\
\affaddr{Cornell University}\\
\affaddr{Ithaca, NY 14853-7501}\\
\email{kurland@cs.cornell.edu, llee@cs.cornell.edu}
}
\conferenceinfo{SIGIR'06,} {August 6--11, 2006, Seattle, Washington, USA.} 
\CopyrightYear{2006} 
\crdata{1-59593-369-7/06/0008} 

\newcommand{\query}{q}
\newcommand{\doc}{d}
\newcommand{\clust}{c}
\newcommand{\titem}{s}

\newcommand{\wordseqvar}{s} 

\newcommand{\altArbGroupMember}{x}

\newcommand{\set}[1]{\{#1\}}
\newcommand{\definedas}{\stackrel{def}{=}}

\newcommand{\kld}[2]{D\left(#1 \; \Big\vert\Big\vert \,\, #2\right)}

\newcommand{\relret}[1]{}

\definecolor{lightgray}{gray}{.75}
\providecommand{\epsfscaledbox}[2]{\centerline{\psfig{figure=#1,width=#2}}}

\newcommand{\prob}{p}

\newcommand{\ilmprob}{\prob} 

\newcommand{\inducedprob}[3]{\ensuremath{#1_{#2}(#3)}}

\newcommand{\mlprobTerm}[2]{\inducedprob{\widetilde{\ilmprob}^{\,MLE}}{#1}{#2}}

\newcommand{\dirichletParam}{\mu}
\newcommand{\dirichletParamEvalSpecific}{\mu(\evaluationMeasure)}
\newcommand{\dirichletLM}[3]{\inducedprob{\ilmprob^{[\dirichletParam]}}{#1}{#2}}
\newcommand{\dirichletLMTerm}[3]{\dirichletLM{#1}{#2}{#3}}

\newcommand{\numDocsInGroup}{N}

\newcommand{\initTag}{{\rm init}}
\newcommand{\topRetGroup}{{\cal D}_{\initTag}}
\newcommand{\clustSetVar}{{\mathit Cl(\topRetGroup)}}
\newcommand{\initSet}{{\cal S}_{\initTag}}

\newcommand{\evaluationMeasure}{e} 

\newcommand{\prestigeText}{centrality\xspace}
\newcommand{\PrestigeText}{Centrality\xspace}

\newcommand{\prestigious}{central\xspace}

\newcommand{\pageRank}{{\sf PR}\xspace}

\newcommand{\pointTo}{\to}

\newcommand{\dampFactor}{\lambda}

\newcommand{\arbGraph}{G}
\newcommand{\arbGraphVertices}{V}

\newcommand{\dToD}{document-to-document\xspace}
\newcommand{\dToDAbbrev}{\mbox{d$\leftrightarrow$d}\xspace}

\newcommand{\edgeAssert}[2]{#1 \pointTo #2}
\newcommand{\edge}[2]{(\edgeAssert{#1}{#2})}
\newcommand{\wtName}{{\mathit wt}}
\newcommand{\edgeWeight}[2]{\wtName\edge{#1}{#2}}

\newcommand{\outwt}[1]{out(#1)}

\newcommand{\node}{v}
\newcommand{\nodeFrom}{u}
\newcommand{\nodeTo}{v}

\newcommand{\wtedInDegree}{influx\xspace}
\newcommand{\WtedInDegree}{Influx\xspace}
\newcommand{\supportGraph}{relevance-flow graph\xspace}
\newcommand{\supportGraphs}{relevance-flow graphs\xspace}

\newcommand{\hubvar}{{\sf hub}}

\newcommand{\authvar}{{\sf auth}}

\newcommand{\rightbip}{one-way bipartite\xspace}
\newcommand{\leftside}{\arbGraphVertices_{\rm Left}}
\newcommand{\rightside}{\arbGraphVertices_{\rm Right}}

\newcommand{\numGenerators}{\delta}

\newcommand{\cToD}{document-as-authority\xspace}
\newcommand{\cToDAbbrev}{\mbox{c{$\rightarrow$}d}\xspace}

\newcommand{\dToC}{document-as-hub\xspace}
\newcommand{\dToCAbbrev}{\mbox{d{$\rightarrow$}c}\xspace}

\newcommand{\generators}{{\mathit Nbhd}}

\newcommand{\clustSize}{k}

\newcommand{\paramGenGroupOfItem}[3]{\generators(#1\,\vert\,#3,#2)} 

\newcommand{\arbNumGenerators}{m}

\newcommand{\firstmention}[1]{{\bf #1}}

\newcommand{\docText}{Doc}
\newcommand{\clustText}{Clust}

\newcommand{\HITS}{HITS\xspace}

\newcommand{\abbrevInit}{init. ranking}

\newcommand{\rerankApproach}{structural \mbox{re-rank}\-ing\xspace}

\newcommand{\reranking}{re-rank\-ing\xspace}
\newcommand{\ReRanking}{Re-Ranking\xspace}
\newcommand{\rerank}{re-rank\xspace}

\newcommand{\RERANKING}{RE-RANKING\xspace}

\newcommand{\statSymbolInit}{i}
\newcommand{\statSymbolOpt}{o}
\newcommand{\statSymbolPR}{p}
\newcommand{\statSymbolHITS}{h}
\newcommand{\statSymbolCQL}{c}

\newcommand{\authSymbol}{A}
\newcommand{\hubSymbol}{H}

\newcommand{\statSymbolCompTables}{*}

\newcommand{\smoothWtName}[1]{\wtName^{[#1]}}
\newcommand{\edgeWeightSmooth}[3]{\smoothWtName{#3}\edge{#1}{#2}}
\newcommand{\prAlgorithm}{PageRank\xspace}

\newcommand{\bipText}{\rightbip}

\newcommand{\docBasedRetrieval}{documents?}
\newcommand{\clustBasedRetrieval}{clusters?}
\newcommand{\abbrevCQL}{clust-$\dirichletLMTerm{\clust}{\query}{}$\xspace}
\newcommand{\veryAbbrevCQL}{$\dirichletLMTerm{\clust}{\query}{}$}
\newcommand{\abbrevAuth}{Auth}
\newcommand{\abbrevHub}{Hub}
\newcommand{\abbrevPR}{PageRank}
\newcommand{\abbrevInflux}{Influx}
\newcommand{\abbrevDoc}{doc}
\newcommand{\abbrevClust}{clust}
\newcommand{\abbrevDocAuthText}{\abbrevDoc-\abbrevAuth\xspace}

\newcommand{\abbrevDocHub}[1]{\abbrevDoc-\abbrevHub({#1})}
\newcommand{\abbrevDocInfluxText}{\abbrevDoc-\abbrevInflux\xspace}
\newcommand{\abbrevDocPRText}{\abbrevDoc-\abbrevPR\xspace}

\newcommand{\abbrevDocInflux}[1]{\abbrevDoc-\abbrevInflux({#1})}
\newcommand{\abbrevDocPR}[1]{\abbrevDoc-\abbrevPR({#1})}
\newcommand{\abbrevClustAuth}[1]{\abbrevClust-\abbrevAuth({#1})}
\newcommand{\abbrevClustInflux}[1]{\abbrevClust-\abbrevInflux({#1})}
\newcommand{\abbrevClustPR}[1]{\abbrevClust-\abbrevPR({#1})}
\newcommand{\abbrevDocAuthDtoD}{\abbrevDocAuth{\dToDAbbrev}}

\newcommand{\abbrevDocPRdTod}{\abbrevDocPR{\dToDAbbrev}}
\newcommand{\abbrevDocAuthCtoD}{\abbrevDocAuth{\cToDAbbrev}}

\newcommand{\abbrevDocInfluxCtoD}{\abbrevDocInflux{\cToDAbbrev}}
\newcommand{\abbrevDocPRcToD}{\abbrevDocPR{\cToDAbbrev}}
\newcommand{\abbrevClustAuthDtoC}{\abbrevClustAuth{\dToCAbbrev}}
\newcommand{\veryAbbrevClustAuthDtoC}{\abbrevAuth[{\dToCAbbrev}]}

\newcommand{\abbrevClustInfluxDtoC}{\abbrevClustInflux{\dToCAbbrev}}
\newcommand{\veryAbbrevClustInfluxDtoC}{\abbrevInflux[{\dToCAbbrev}]}
\newcommand{\abbrevClustPRdToC}{\abbrevClustPR{\dToCAbbrev}}

\begin{document}
\maketitle

\begin{abstract}
We present an approach to improving the precision of an initial
document ranking wherein we utilize cluster information within a
graph-based framework.  The main idea is to perform re-rank\-ing based
on centrality within bipartite graphs of documents (on one side) and
clusters (on the other side), on the premise that these are mutually
reinforcing entities.  Links between entities are created via
consideration of language models induced from them.

We find that our cluster-document graphs give rise to much better
retrieval performance than previously proposed document-only graphs
do.  For example, authority-based re-rank\-ing of documents via a
HITS-style cluster-based approach outperforms a previously-proposed
PageRank-inspired algorithm applied to solely-document graphs.
Moreover, we also show that computing authority scores for clusters
constitutes an effective method for identifying clusters containing a
large percentage of relevant documents.
\end{abstract}

\vspace{1mm}
{
\small
\noindent
{\bf Categories and Subject Descriptors:} H.3.3 {[Information Search
and Retrieval]}: {Retrieval models}

\vspace{1mm}
\noindent
{\bf General Terms:} Algorithms, Experimentation

\vspace{1mm}
}
\section{Introduction}
\label{sec:intro}

To improve the precision of  retrieval output, 
especially within the very few (e.g, 5 or 10) highest-ranked documents
that are returned, a number of researchers
\cite{Willett:85a,Hearst+Pedersen:96a,Kleinberg:98a,Danilowicz+Balinski:01a,Leuski:01a,Tombros+Villa+Rijsbergen:02a,Liu+Croft:04a,Balinski+Danilowicz:05a,Kurland+Lee:05a,Diaz:05a}
have considered a {\em \rerankApproach} strategy.
The
idea is to \rerank the top $\numDocsInGroup$ documents that some initial search
engine produces, where the re-ordering utilizes
information about inter-document relationships within that set.
Promising results have been previously obtained by using document {\em
\prestigeText} within the initially retrieved list to perform
\rerankApproach, on the premise that if the quality of this list is
reasonable to begin with, then the documents that are most related to most
of the documents on the list are likely to be the most relevant ones.  In
particular, in our prior work \cite{Kurland+Lee:05a} we adapted {\em
PageRank} \cite{Brin+Page:98a} --- which, due to
the success of Google,  is surely the most well-established 
algorithm
for defining and computing \prestigeText within a directed graph ---  
to the task of \reranking non-hyperlinked document sets.

The arguably most well-known alternative to PageRank is Kleinberg's
{\em \HITS} algorithm \cite{Kleinberg:98a}.  The major conceptual way
in which \HITS differs from PageRank is that it defines two different
types of \prestigious items:
each node
is assigned
both a {\em hub} and an {\em
authority} score 
as opposed to a single PageRank score. In the Web
setting, in which \HITS was originally proposed, good hubs correspond
roughly to high-quality resource lists or collections of pointers,
whereas good authorities correspond to the high-quality resources
themselves; thus, distinguishing between two differing but
interdependent types of Webpages is quite appropriate.  
Our previous study \cite{Kurland+Lee:05a}
applied HITS
to non-Web documents.
We found that its performance was comparable
to or better than that of algorithms that do not involve
\rerankApproach; however, HITS was not as effective as 
PageRank \cite{Kurland+Lee:05a}.

Do these results imply that PageRank is better than \HITS for \rerankApproach of
non-Web documents?  Not
necessarily, because there may exist graph-construction methods that
are more suitable for \HITS.  Note that the only entities considered in
our previous study were documents. If we could introduce entities
distinct from documents but enjoying a mutually reinforcing
relationship with them, then we might better satisfy the spirit of the
hubs-versus-authorities distinction, and thus derive stronger results
utilizing \HITS.

A crucial insight of the present paper is that document {\em clusters}
appear extremely well-suited to play this complementary role.  The
intuition is that: (a) given those clusters that are 
``most
representative'' of the user's information need, the documents within
those clusters are likely to be relevant; and (b) the ``most
representative'' clusters 
should be those that contain many relevant
documents.      This apparently circular reasoning is strongly
reminiscent of
   the inter-related hubs and authorities concepts underlying \HITS.  

Also, clusters have long been considered a promising source of
information.  The well-known {\em cluster hypothesis}
\cite{vanRijsbergen:79a} 
encapsulates the intuition that clusters can  reveal groups of
relevant documents; in practice, the potential utility of clustering
for this purpose has been demonstrated 
for both the case wherein clusters were created in a query-independent
fashion \cite{Jardine+Rijsbergen:71a,Croft:80a} and
the \reranking setting
\cite{Hearst+Pedersen:96a,Leuski:01a,Tombros+Villa+Rijsbergen:02a}.

In this paper, we show through an array of experiments that
consideration of the mutual reinforcement of clusters and documents in
determining \prestigeText can lead to highly effective algorithms for
\reranking an initially retrieved list. 
Specifically, our experimental results show that 
the
\prestigeText-induction methods
that we previously studied solely in the context of document-only
graphs \cite{Kurland+Lee:05a}
result in 
much better \reranking performance if implemented
over bipartite graphs of documents (on one side) and clusters (on
the other side). For example,
ranking 
{\em documents} by their ``authoritativeness'' as computed by HITS upon these 
cluster-document graphs yields better performance than
that of a previously proposed PageRank implementation 
applied to
document-only graphs.
Interestingly, we also find that 
{\em cluster} authority scores can be used
to identify clusters containing a large percentage of relevant
documents.

\newtheorem{theorem}{Theorem}
\newcommand{\pageRankBip}{\pageRank_{\mathit bip}}
\renewcommand{\docText}{doc}
\renewcommand{\clustText}{clust}
\newcommand{\generalNode}{n}

\section{Algorithms for \RERANKING}
\label{sec:models}

Since we are focused on the \rerankApproach paradigm, our algorithms
are applied not to the entire corpus, but to a
subset 
$\topRetGroup^{\numDocsInGroup,\query}%
$ (henceforth $\topRetGroup$),  defined as the
top $\numDocsInGroup$ documents retrieved in response to the query $\query$
by a given initial retrieval engine. Some of our algorithms
also take into account a set $\clustSetVar%
$ of {\em clusters} of
the documents in $\topRetGroup$.  We use $\initSet$ to refer
generically to
whichever set of entities --- either $\topRetGroup$ or $\topRetGroup
\cup \clustSetVar$ --- is used by a given algorithm.

The basic idea behind the algorithms we consider is to determine
\prestigeText within a 
{\em \supportGraph}, defined as a directed graph with non-negative weights on the
edges in which 
\begin{itemize}
\item the nodes are the elements of $\initSet$,
and 
\item
 the weight on an edge between node $\nodeFrom$ and $\nodeTo$
  is based on the strength of
  evidence for $\nodeTo$'s  relevance  
that would follow from an assertion that  $\nodeFrom$ is relevant.
\end{itemize}
By construction, then, any measure of the \prestigeText of 
$\titem \in \initSet$ should measure
the accumulation of evidence for its relevance according to the 
set of interconnections among  the entities in $\initSet$.  Such information can
then optionally be subjected to additional processing, such as
integration with information on each item's similarity to the query,
to produce a final \reranking of $\topRetGroup$.

\paragraph*{Conventions regarding graphs} The types of \supportGraphs we
consider can all be represented as weighted directed graphs of the
form $(\arbGraphVertices,\wtName)$, where $\arbGraphVertices$ is a
finite non-empty set of nodes and 
$\wtName: \arbGraphVertices \times \arbGraphVertices \rightarrow [0,\infty)$ is a non-negative edge-weight function.
Note that thus our graphs technically have edges between all ordered pairs of
nodes (self-loops included);  however, edges with zero
edge-weight are conceptually equivalent to missing edges.  
For clarity, we write
$\edgeWeight{\nodeFrom}{\nodeTo}$ instead of
$\wtName(\nodeFrom,\nodeTo)$.

\subsection{Hubs, authorities, and the \HITS algorithm}  
\label{sec:hits}

The \HITS
algorithm for computing \prestigeText can be motivated as follows.
Let $\arbGraph = (\arbGraphVertices,\wtName)$ be the input graph, and let $\nodeTo$ be
a node in $\arbGraphVertices$.  First, suppose 
we somehow knew the {\em
  hub score}  $\hubvar(\nodeFrom)$ of each node $\nodeFrom \in \arbGraphVertices$, where
``hubness'' is the extent to which the nodes that $\nodeFrom$ points
to are ``good'' in some sense. Then, $\nodeTo$'s {\em authority score} 
\begin{equation}
\authvar(\nodeTo) = \sum_{\nodeFrom \in \arbGraphVertices}
\edgeWeight{\nodeFrom}{\nodeTo} \cdot \hubvar(\nodeFrom) 
\label{eq:auth}
\end{equation}
would be a natural measure of how ``good'' $\node$ is, since a node that is
``strongly'' pointed to by high-quality hubs (which, by definition, tend to point to
``good'' nodes) receives a high score.
But where do we get the hub score
for 
a given node
$\nodeFrom$?  A natural choice is to use the extent to which $\nodeFrom$
``strongly'' points to highly  authoritative nodes:
\begin{equation}
\hubvar(\nodeFrom) = \sum_{\nodeTo \in \arbGraphVertices}
 \edgeWeight{\nodeFrom}{\nodeTo}\cdot \authvar(\nodeTo).
\label{eq:hub}
\end{equation}
Clearly, Equations \ref{eq:auth} and \ref{eq:hub} are mutually
recursive
. However, 
the iterative HITS algorithm%
\footnote{Strictly speaking, the algorithm and proof of convergence as originally
  presented \cite{Kleinberg:98a} need (trivial) modification to apply to 
edge-weighted graphs.} 
provably converges to 
(non-identically-zero, non-negative) score functions $\hubvar^*$ and $\authvar^*$
that satisfy 
the above pair of equations
.  

\begin{figure}[ht]
\epsfscaledbox{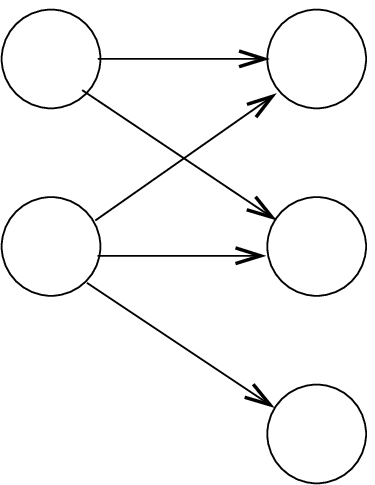}{.5in}
\caption{\label{fig:hits-icon} A \rightbip graph.  We only show positive-weight
  edges (omitting weight values).  According to HITS, the left-hand
  nodes are (pure) hubs; the right-hand ones are (pure)
  authorities.}
\end{figure}

Figure
\ref{fig:hits-icon} depicts the ``iconic'' case in which the input graph
$\arbGraph$ is  {\em \rightbip}, that is, $\arbGraphVertices$ can be
partitioned into non-empty sets $\leftside$ and
$\rightside$ such that only edges in $\leftside \times \rightside$
can receive positive weight, and $\forall \nodeFrom \in \leftside$,
$\sum_{\nodeTo \in \rightside} \edgeWeight{\nodeFrom}{\nodeTo} > 0$.
It is the case that $\authvar^*(\nodeFrom) = 0$
for every $\nodeFrom \in \leftside$ and $\hubvar^*(\nodeTo)=0$ for
every $\nodeTo \in \rightside$; in this sense, the left-hand
nodes are ``pure'' hubs and the right-hand nodes are ``pure''
authorities.

Note that in the end, we need to produce a {\em single}
\prestigeText score 
for each node $\generalNode \in \arbGraphVertices$%
.  For experimental simplicity, we consider only two possibilities in
this paper --- 
using $\authvar^*(\generalNode)$
as the final \prestigeText score, 
or using 
$\hubvar^*(\generalNode)$ instead%
 --- although combining the hub and authority scores is also an
    interesting possibility.

\subsection{Graph schemata: incorporating clusters}
\label{sec:schemata}

Recall that the fundamental operation in our \rerankApproach paradigm
is to compute the \prestigeText of entities (with)in
a set $\initSet$.
One possibility is to define $\initSet$ as $\topRetGroup$, the documents in the
initially retrieved set; we refer
generically to any \supportGraph induced under this choice as a \firstmention{{\em
\dToD}}  graph.  But note that  for non-Web documents, it may not be
obvious {\em a priori} what kinds of documents are hubs and what kinds are
authorities.

Alternatively, we can 
define
$\initSet$ as $\topRetGroup \cup \clustSetVar$, where $\clustSetVar$
consists of clusters of the documents in $\topRetGroup$.
On a purely formal level, doing so allows us to
map the hubs/\hfill authorities  duality
discussed above 
onto the
documents/clusters duality, as follows.  Recalling our discussion of
the ``iconic'' case of \rightbip graphs
$\arbGraph=((\leftside,\rightside),\wtName)$,
we can create \firstmention{{\em \cToD}} graphs 
simply by choosing
$\leftside=\clustSetVar$ and
$\rightside=\topRetGroup$,
so that
necessarily 
clusters serve the role of (pure) hubs and 
documents serve the role of (pure) authorities.
Contrariwise,\footnote{In practice, one can
simultaneously compute the 
output of \HITS
for a given \cToD and \dToC
graph {\em pair} by ``overlaying'' the two into a single graph and suitably
modifying \HITS's normalization scheme.}  we can create
\firstmention{{\em \dToC}} graphs by setting $\leftside=\topRetGroup$
and $\rightside=\clustSetVar$.

But the advantages of incorporating cluster-based information are
not just formal.  The well-known {\em cluster hypothesis}
\cite{vanRijsbergen:79a} 
encapsulates the intuition that clusters can reveal groups of
relevant documents; in practice, the potential utility of clustering
for this purpose has been demonstrated a number of 
times,
whether 
the
clusters were created in a query-independent fashion
\cite{Jardine+Rijsbergen:71a,Croft:80a}, or from
the initially most-highly-ranked documents for some query
\cite{Hearst+Pedersen:96a,Leuski:01a,Tombros+Villa+Rijsbergen:02a}
(i.e., in 
the \reranking setting).  Since \prestigious clusters are,
supposedly, those that accrue the most evidence for relevance,
documents that are strongly identified with such clusters should
themselves be judged highly relevant.\footnote{We say ``are strongly identified with'', as opposed to
``belong to'' to allow for overlapping or probabilistic clusters.
Indeed, the \rightbip graphs we construct are ill-suited to the HITS
algorithm if document-to-cluster links are based on membership in
disjoint clusters.}
\footnote{This is, in some sense, a type of smoothing: a
document might be missing some of the query terms (perhaps due to synonymy), but if it lies
within a sector of ``document space'' containing many relevant documents, it could still be
deemed highly relevant.  
Recent research pursues this smoothing
idea at a deeper level \cite{Liu+Croft:04a,Kurland+Lee:04a}.
}
But identifying such clusters is facilitated by knowledge of which
documents are most likely to be relevant --- exactly the mutual
reinforcement property that \HITS was designed to leverage.

\subsection{Alternative scores: PageRank and \wtedInDegree}
\label{sec:pagerank}

We will compare the results of using the \HITS algorithm 
against those derived using PageRank instead.  This is a natural 
comparison because PageRank is the most well-known
\prestigeText-induction algorithm utilized for ranking documents, and 
because in earlier work \cite{Kurland+Lee:05a}, PageRank performed
quite well as a tool for \rerankApproach of non-Web documents,  at
least when applied to \dToD graphs.

One can think of PageRank as a version of \HITS in which the
hub/authority distinction has been collapsed.
Thus, writing ``\pageRank'' for both $\authvar$ and
$\hubvar$, we {\em conceptually} have the (single) equation
\begin{equation}
{\pageRank}(\nodeTo) = \sum_{\nodeFrom \in \arbGraphVertices}
\edgeWeight{\nodeFrom}{\nodeTo} \cdot {\pageRank}(\nodeFrom).
\label{eq:pagerank-concept}
\end{equation}
However, in practice, we incorporate Brin and
Page's smoothing scheme
\cite{Brin+Page:98a} 
together with a 
correction for nodes with no positive-weight edges emanating from them
\cite{Ng+Zheng+Jordan:01a,Langville+Meyer:05a}:
\begin{eqnarray}
\pageRank(\nodeTo) & = & \sum_{\nodeFrom \in \arbGraphVertices:  \outwt{\nodeFrom}
  > 0} \left[\frac{(1 - \dampFactor)}{|\arbGraphVertices|} +
\dampFactor
\frac{\edgeWeight{\nodeFrom}{\nodeTo}}{\outwt{\nodeFrom}}\right]\cdot
\pageRank(\nodeFrom) \nonumber
\\ 
& +&  \sum_{\nodeFrom \in \arbGraphVertices: \outwt{\nodeFrom} = 0} \frac{1}{|\arbGraphVertices|}\cdot \pageRank(\nodeFrom)
\label{eq:pagerank}
\end{eqnarray}
where  $\outwt{\nodeFrom} \definedas \sum_{\nodeTo' \in
\arbGraphVertices}\edgeWeight{\nodeFrom}{\nodeTo'}$, and 
$\lambda \in (0,1)$ is the damping factor.\footnote{Under the
original
``random surfer'' model, the sum of the transition probabilities out
of ``no outflow'' nodes --- which are abundant in \rightbip graphs ---
 would be $(1 - \dampFactor)$, not 1.  Conceptually, 
the role of the second summation  in Equation \ref{eq:pagerank} is to set
$\dampFactor=0$ for these no-outflow nodes.
}

Equation \ref{eq:pagerank} is recursive, but there are 
iterative algorithms that provably converge to the unique positive 
solution  $\pageRank^*$ satisfying the sum-normalization constraint $\sum_{\node \in \arbGraphVertices} \pageRank(\node)=1$
\cite{Langville+Meyer:05a}.  
Moreover, a (non-trivial) {\em closed-form} --- and quite easily
 computed --- 
 solution exists for \rightbip graphs:
\begin{theorem}
If $\arbGraph=(\arbGraphVertices,\wtName)$ is \rightbip, then 
\begin{equation}
\pageRankBip(\nodeTo) \definedas
\sum_{\nodeFrom \in \arbGraphVertices: \\ \outwt{\nodeFrom}  > 0}
  \frac{\edgeWeight{\nodeFrom}{\nodeTo}}{\outwt{\nodeFrom}}
\label{eq:bip-pagerank}
\end{equation}
is an affine transformation (with respect to positive constants) of, and therefore equivalent for
ranking purposes to, the unique positive sum-normalized solution to
Equation \ref{eq:pagerank}.  
\label{thm:pagerankBip}
\end{theorem} 
\vspace{-0.05in}
(Proof omitted due to space constraints.)
Interestingly, this result shows that while one might have thought
that clusters and documents would ``compete'' for PageRank score when
placed within the same graph, in our \cToD and \dToC graphs this is not the case.

Earlier work \cite{Kurland+Lee:05a} also considered scoring a node
$\nodeTo$ by its {\em \wtedInDegree},  $\sum_{\nodeFrom \in \arbGraphVertices}
\edgeWeight{\nodeFrom}{\nodeTo}$.  This can be viewed as either a
non-recursive version of Equation \ref{eq:pagerank-concept}, or as an
un-normalized analog 
of Equation \ref{eq:bip-pagerank}.

\subsection{Algorithms based on \prestigeText scores}
\label{sec:algNames}

Clearly, we can rank documents by their scores as computed by
any of the functions introduced above.  But when we operate on \cToD
or \dToC graphs, \prestigeText scores for the clusters are also
produced.  These can be used to derive alternative
means for ranking documents.  We follow 
Liu and Croft's approach \cite{Liu+Croft:04a}: first, 
rank the documents within (or most strongly associated to) each cluster
according to the initial retrieval engine's scores; then, derive the
final list by concatenating the within-cluster lists in order of
decreasing cluster score, discarding repeats.  Such an approach would
be successful if 
cluster  \prestigeText is strongly correlated with
the property of containing a
large
percentage of relevant documents.

\paragraph*{Ranking algorithms}
Since we have two possible ranking para\-digms, we adopt the following
{\bf algorithm  naming conventions}.
Names consist of a hyphen-separated prefix and suffix. The prefix
(``\docText'' or ``\clustText'') indicates whether documents
were ranked directly by their \prestigeText scores, or indirectly through the
concatenation process outlined above in which it is the {\em clusters'}
\prestigeText scores that were employed.  The suffix (``Auth'',
``Hub'', ``\pageRank'', or ``\WtedInDegree'') indicates which score
function ($\authvar^*$,
$\hubvar^*$, $\pageRank^*$ (or $\pageRankBip$), or \wtedInDegree) was
used to measure \prestigeText.  
For a given \reranking algorithm, we indicate the graph upon which it
was run in brackets, e.g., ``doc-Auth[$G$]''.

\section{Related Work}
\label{sec:relwork}

The potential merits of {\em query-dependent clustering}, that is,
 clustering
the documents 
retrieved in response to a query,
have long been
recognized \cite{Preece:73a,Willett:85a,Leuski+Allan:98a,Tombros+Villa+Rijsbergen:02a,Liu+Croft:04a},
 especially in interactive retrieval settings
\cite{Hearst+Pedersen:96a,Leuski:01a,Shen+Zhai:05a}. However, automatically
detecting clusters that contain many relevant documents remains a very
hard task \cite{Willett:85a}.
Section
\ref{sec:clustAuth} presents results for detecting such clusters using
\prestigeText-based cluster ranking.

Recently, there has been a growing body of work on graph-based
modeling for different language-processing 
tasks where\-in links are induced by inter-entity
textual similarities. Examples include 
document 
\mbox{(re-)ranking}
\cite{Danilowicz+Balinski:01a,Levow+Matveeva:04a,Diaz:05a,Kurland+Lee:05a,Zhang+al:05a},
text summarization
\cite{Erkan+Radev:04b,Mihalcea+Tarau:04a}, 
sentence retrieval 
\cite{Otterbacher+Erkan+Radev:05a}, and document representation \cite{Erkan:06a}. In contrast
to our methods,
links connect entities of the same type, and
clusters of entities are not modeled within the graphs.

While ideas similar to ours by 
virtue of leveraging the mutual
reinforcement 
of entities of different 
types, or using
bipartite graphs of such entities for clustering (rather than using
clusters), are abundant (e.g.,
\cite{Karov+Edelman:98a,Dhillon:01a,Beeferman+Berger:00a}), we focus
here on exploiting mutual reinforcement in ad hoc retrieval.

Random walks (with early stopping) over bipartite graphs of
terms and documents were used for query expansion
\cite{Lafferty+Zhai:01a}, but 
in contrast to our work, no
stationary solution 
was sought. A similar ``short chain''
approach utilizing bipartite graphs of clusters and documents for
ranking an entire corpus was recently proposed
\cite{Kurland+Lee+Domshlak:05a}, thereby constituting the work most
resembling ours. However, 
again, a stationary distribution was not sought. Also,
{\em query drift} prevention mechanisms were 
required to obtain good
performance; in our \reranking setting, we 
need not employ such mechanisms.

\section{Evaluation Framework}
\label{sec:eval}

\newcommand{\abbrevDocAuth}[1]{\abbrevDoc-\abbrevAuth[{#1}]\xspace}
\renewcommand{\abbrevDocHub}[1]{\abbrevDoc-\abbrevHub[{#1}]\xspace}
\renewcommand{\abbrevDocInflux}[1]{\abbrevDoc-\abbrevInflux[{#1}]\xspace}
\renewcommand{\abbrevDocPR}[1]{\abbrevDoc-\abbrevPR[{#1}]\xspace}
\renewcommand{\abbrevClustAuth}[1]{\abbrevClust-\abbrevAuth[{#1}]\xspace}
\newcommand{\abbrevClustHub}[1]{\abbrevClust-\abbrevHub[{#1}]\xspace}
\renewcommand{\abbrevClustInflux}[1]{\abbrevClust-\abbrevInflux[{#1}]\xspace}
\renewcommand{\abbrevClustPR}[1]{\abbrevClust-\abbrevPR[{#1}]\xspace}
\newcommand{\abbrevDocHubDtoC}{\abbrevDocHub{\dToCAbbrev}}
\newcommand{\abbrevClustHubCtoD}{\abbrevClustHub{\cToDAbbrev}}
\newcommand{\dToDAbbrevSmooth}{d$\leftrightarrow$d[$\dampFactor$]}
\newcommand{\cToDAbbrevSmooth}{d:auth[$\dampFactor$]\xspace}
\newcommand{\abbrevDocAuthDtoDSmooth}{\abbrevDocAuth{\dToDAbbrevSmooth}}
\newcommand{\abbrevDocAuthCtoDSmooth}{\abbrevDocAuth{\cToDAbbrevSmooth}}
\renewcommand{\statSymbolHITS}{a}
\newcommand{\abbrevDocHubText}{\abbrevDoc-\abbrevHub\xspace}
\newcommand{\centScore}[2]{Cen(#1;#2)}
\newcommand{\dToDAbbrevText}{\dToDAbbrev}
\newcommand{\paramEdgeWeight}[3]{\wtName^{[#3]}\edge{#1}{#2}}
\newcommand{\withSmooth}{S}
\newcommand{\withoutSmooth}{U}

\newcommand{\paramEdgeWt}[3]{\wtName^{[#3]}\edge{#1}{#2}} 
\newcommand{\better}[1]{\mathit{#1}}

Most aspects of the evaluation framework described below are
adopted from our previous experiments  with non-cluster-based \rerankApproach \cite{Kurland+Lee:05a}
so as to facilitate direct comparison.  
Section 4.1 of
\cite{Kurland+Lee:05a} provides a more detailed
justification of the experimental
design.  The
main conceptual changes
\footnote{Some of the \prAlgorithm results
appearing in our previous paper \cite{Kurland+Lee:05a} accidentally reflect
experiments utilizing a suboptimal choice of $\topRetGroup$.  For
citation purposes, the numbers reported in the current paper should be
used.} here are: a slightly larger parameter search-space for the
``out-degree'' parameter
$\numGenerators$ (called the ``ancestry'' parameter $\alpha$ in \cite{Kurland+Lee:05a});  and, of course, the incorporation of clusters.

\newcommand{\summaryTableSize}{\small}
\newcommand{\summaryTableCaption}{\label{tab:summaryTable} 
Main comparison:  \HITS 
or
\prAlgorithm on document-only
graphs 
versus 
\HITS on cluster-to-document graphs.
Bold: best results per column.  
Symbols ``$\statSymbolPR$'' and ``$\statSymbolHITS$'': 
\abbrevDocAuthCtoD  result differs 
significantly from that of \abbrevDocPRdTod or \abbrevDocAuthDtoD, respectively.
}

\begin{table*}[t]
\begin{center}
\summaryTableSize
\begin{tabular}{|l||c|c|c|c|c|c|c|c|c|}
\hline
& \multicolumn{3}{|c|}{AP}& \multicolumn{3}{|c|}{TREC8}& \multicolumn{3}{|c|}{WSJ}\\ \cline{2-10} 
 & {prec$@5$} & {prec$@10$} & {MRR} & {prec$@5$} & {prec$@10$} & {MRR} & {prec$@5$} & {prec$@10$} & {MRR} \\ \hline
\abbrevDocAuthDtoD & $ .509 $ $ ^{}_{}$& $ .486 $ $^{}_{}$& $ .638 $ $^{}_{}$& $ .440 $ $ ^{}_{}$& $ .424 $ $^{}_{}$& $ .648 $ $^{}_{}$& $ .504 $ $ ^{}_{}$& $ .464 $ $^{}_{}$& $ .638 $ $^{}_{}$\\ \hline
\abbrevDocPRdTod & $ .519 $ $ ^{}_{}$& $ .480 $ $^{}_{}$& $ .632 $ $^{}_{}$& $ .524 $ $ ^{}_{}$& $ .446 $ $^{}_{}$& $ .666 $ $^{}_{}$& $ .536 $ $ ^{}_{}$& $ .486 $ $^{}_{}$& $ .699 $ $^{}_{}$\\ \hline
\abbrevDocAuthCtoD & \mbox{\boldmath$ .541 $} $^{}_{}$& \mbox{\boldmath$ .501 $} $^{\statSymbolPR}_{}$& \mbox{\boldmath$ .669 $} $^{\statSymbolPR}_{}$& \mbox{\boldmath$ .544 $} $^{\statSymbolHITS}_{}$& \mbox{\boldmath$ .452 $} $^{}_{}$& \mbox{\boldmath$ .674 $} $^{}_{}$& \mbox{\boldmath$ .564 $} $^{\statSymbolHITS}_{}$& \mbox{\boldmath$ .514 $} $^{\statSymbolHITS}_{}$& \mbox{\boldmath$ .746 $} $^{\statSymbolHITS}_{}$\\ \hline
\end{tabular}

\caption{\summaryTableCaption}
\end{center}
\end{table*}

\subsection{Graph construction}
\label{sec:lm}

\paragraph*{Relevance flow based on language models (LMs)}
\newcommand{\simFn}{\mbox{\it rflow}}

\newcommand{\genprobKL}[2]{\simFn(#1,#2)}
\renewcommand{\mlprobTerm}[2]{\inducedprob{{\ilmprob}^{\;[0]}}{#1}{#2}}
\newcommand{\arbseq}{x}
\renewcommand{\wordseqvar}{y}
\newcommand{\support}{relevance flow\xspace}
\newcommand{\supportAdj}{relevance-flow\xspace}
\newcommand{\restrictSet}{R}
 
To estimate the degree to which one item, if considered
relevant, can vouch for the relevance of another, we follow our previous work on document-based graphs \cite{Kurland+Lee:05a} and utilize
$\dirichletLMTerm{\doc}{\cdot}{}$, the unigram {Dirichlet-smoothed}
language
model
induced from a given document
$\doc$  ($\dirichletParam$ is the smoothing parameter) \cite{Zhai+Lafferty:01a}.
To adapt this estimation scheme to settings
involving  clusters, we derive the language model
$\dirichletLMTerm{\clust}{\cdot}{}$  for a cluster $\clust$  by
treating $\clust$ as the (large) document
formed by concatenating\footnote{Concatenation order is irrelevant for
unigram LMs.} its constituent (or most strongly associated)
documents \cite{Kurland+Lee:04a,Liu+Croft:04a,Kurland+Lee+Domshlak:05a}.

The \supportAdj
measure we use
is essentially a directed similarity
in language-model space:
\newcommand{\simDef}[2]{\exp\left(-\kld{\mlprobTerm{#1}{\cdot}}{\dirichletLMTerm{#2}{\cdot}{\dirichletParam}}\right)}
\begin{equation}
\genprobKL{\altArbGroupMember}{\wordseqvar} \definedas
\simDef{\altArbGroupMember}{\wordseqvar},
\label{eq:KLestimate}
\end{equation}
where $D$ is the Kullback-Leibler divergence.
The asymmetry of this measure corresponds nicely to the intuition that
\support is not symmetric \cite{Kurland+Lee:05a}.
Moreover, this function is somewhat insensitive to large length
differences between the items in question \cite{Kurland+Lee:05a},
which is advantageous when both documents and clusters
(which we treat as very long documents)
are considered.

Previous work \cite{Kurland+Lee:05a,Tao+al:06a}
makes heavy use of the
  idea of nearest neighbors in language-model space. It is therefore
  convenient to introduce the notation
  $\paramGenGroupOfItem{\altArbGroupMember}{\restrictSet}{\arbNumGenerators}$,
  pronounced ``neighborhood'',
to denote the $\arbNumGenerators$ items
  $\wordseqvar$ within the ``restriction set'' $\restrictSet$ that
  have the highest values of
  $\genprobKL{\altArbGroupMember}{\wordseqvar}$ (we break ties by item
  ID, assuming that these have been assigned to documents and
  clusters).
Note that the neighborhood of $\altArbGroupMember$ corresponds to what we
previously termed the ``top generators'' of  $\altArbGroupMember$ \cite{Kurland+Lee:05a}.

\paragraph*{Graphs used in experiments}

For a given set $\topRetGroup$ of initially retrieved documents and
positive integer $\numGenerators$ (an ``out-degree'' parameter), we
consider the following three graphs.
Each connects nodes $\nodeFrom$ to the $\numGenerators$ other nodes,
drawn from some specified set, that  $\nodeFrom$ has the highest
\support to.

The \dToD graph
\firstmention{\dToDAbbrev} has vertex set $\topRetGroup$ and 
weight function $$\wtName^{\dToDAbbrev}(\nodeFrom,\nodeTo) =
\begin{cases} 
\genprobKL{\nodeFrom}{\nodeTo} & \text{if $\nodeTo \in \paramGenGroupOfItem{\nodeFrom}{\topRetGroup-\set{\nodeFrom}}{\numGenerators}$}, \\
 0 & \text{otherwise}.
\end{cases}$$

\newcommand{\ind}{\hspace*{0.1in}}
The \cToD graph \firstmention{\cToDAbbrev} has vertex set
$\topRetGroup
\cup \clustSetVar$  and a weight function such that positive-weight
edges go only from clusters to documents:
$$\wtName^{\cToDAbbrev}(\nodeFrom,\nodeTo) = \\
\begin{cases} 
\genprobKL{\nodeFrom}{\nodeTo} & \text{if $\nodeFrom \in \clustSetVar$
 and} \\
 & \ind \nodeTo \in \paramGenGroupOfItem{\nodeFrom}{\topRetGroup}{\numGenerators}, \\
 0 & \text{otherwise}.
\end{cases}$$

The \dToC graph \firstmention{\dToCAbbrev} has vertex set
$\topRetGroup
\cup \clustSetVar$  and a weight function such that positive-weight
edges go only from documents to clusters:
$$\wtName^{\dToCAbbrev}(\nodeFrom,\nodeTo) = \\
\begin{cases} 
\genprobKL{\nodeFrom}{\nodeTo} & \text{if $\nodeFrom \in \topRetGroup$
 and} \\
 & \ind \nodeTo \in \paramGenGroupOfItem{\nodeFrom}{\clustSetVar}{\numGenerators}, \\
 0 & \text{otherwise}.
\end{cases}$$

Since the latter two graphs are \rightbip, Theorem
\ref{thm:pagerankBip} applies to them.

\paragraph*{Clustering Method} Clearly, our cluster-based graphs require
the construction of clusters of the documents in $\topRetGroup$.
Since this set is query-dependent, at least some of the clustering
process must occur at retrieval time, mandating the use of extremely
efficient algorithms \cite{Cutting+al:92a,Zamir+Etzioni:98a}.  
The approach we adopt is to use
overlapping nearest-neighbor clusters,
which have
formed the basis
of effective retrieval algorithms in other work \cite{Griffiths+Luckhurst+Willett:86a,Kurland+Lee:04a,Kurland+Lee+Domshlak:05a,Tao+al:06a}: 
for each document $\doc \in
\topRetGroup$, we have the cluster
$\set{\doc} \cup \paramGenGroupOfItem{\doc}{\topRetGroup - \set{\doc}}{\clustSize-1}$,
where  $\clustSize$ is the cluster-size parameter.

\subsection{Experimental Setup}

We conducted our experiments on three TREC datasets:
\begin{center}
\begin{tabular}{|lrlc|}\hline
corpus & \# of docs & \multicolumn{1}{c}{queries} & disk(s)  \\ \hline
AP & 242,918 &  51-64, 66-150 & 1-3 \\ \hline
TREC8 & 528,155 & 401-450 & 4-5\\ \hline
WSJ & 173,252 & 151-200 & 1-2  \\ \hline
\end{tabular}
\end{center}
We applied basic tokenization and Porter stemming via the 
Lemur toolkit  (www.lemurproject.org), which we also used  for
language-model induction.
Topic titles served as queries.

In many retrieval situations of interest, ensuring that the top few
documents retrieved (a.k.a., ``the first page of results'') tend to be
relevant is much more important than ensuring that we assign
relatively high ranks to the entire set of
relevant documents in aggregate \cite{Shah+Croft:04a}.  Hence, rather than use mean average precision (MAP) as an
evaluation metric, we apply metrics more appropriate to the
\rerankApproach task:  precision at the top $5$ and $10$
documents (henceforth prec@5 and prec@10, respectively) and the mean reciprocal rank (MRR) of the first relevant
document
\cite{Shah+Croft:04a}. All performance numbers
are averaged over the set of queries for a given corpus.

The natural baseline for the work described here is the standard
language-model-based retrieval approach \cite{Ponte+Croft:98a,Croft+Lafferty:03a}, since it is an effective
paradigm that makes no explicit
use of inter-document relationships.  Specifically, for a given
evaluation metric $\evaluationMeasure$, the corresponding {\em
optimized baseline} is the ranking on documents produced by
$\inducedprob{\ilmprob^{[\dirichletParamEvalSpecific]}}{\doc}{\query}$,
where $\dirichletParamEvalSpecific$ is the value of the Dirichlet
smoothing parameter that results in the best retrieval performance as measured by
$\evaluationMeasure$.

A ranking method might assign different items the same score; we
break such ties by item ID. Alternatively, 
the
scores used to determine
$\topRetGroup$ can be utilized, if available.

{ \newcommand{\graphwidth}{2.8in}
\begin{figure}[h]
\begin{tabular}{c}
\psfig{figure=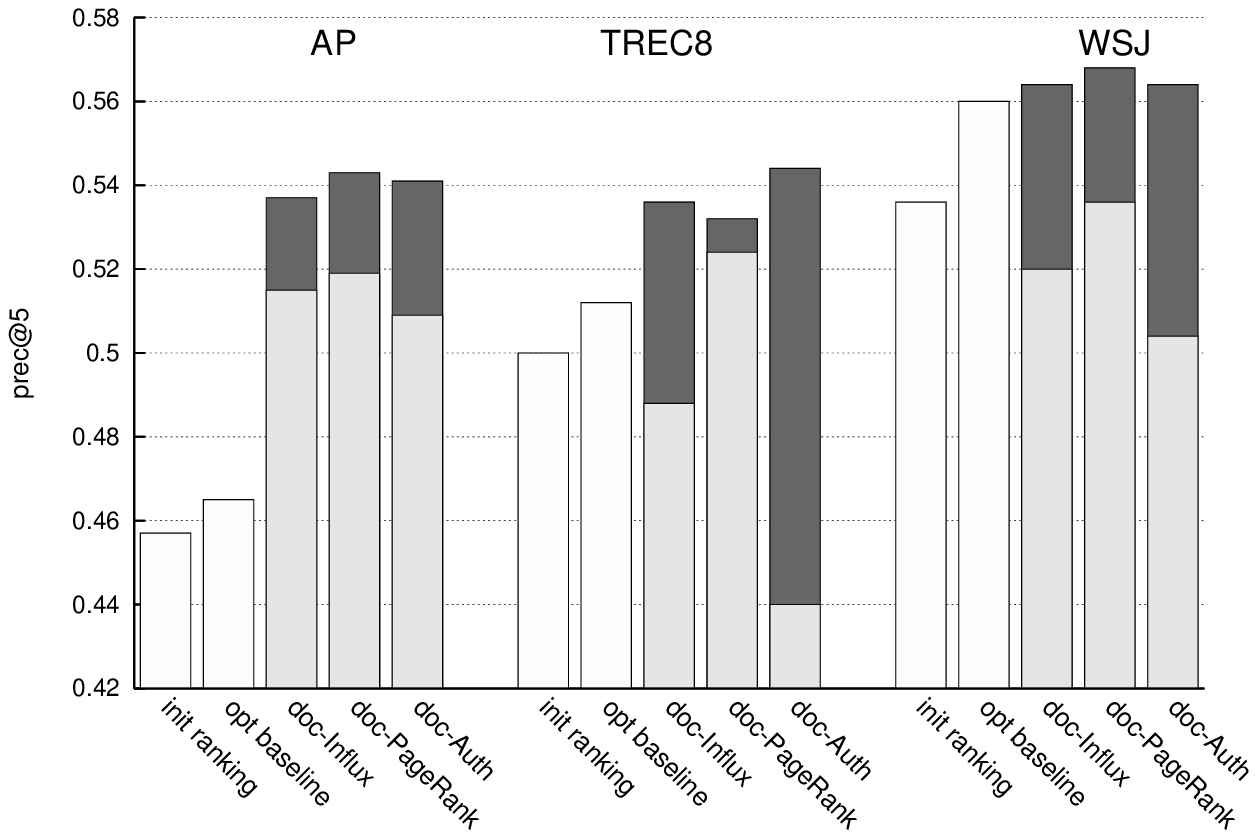,width=\graphwidth} \\
\psfig{figure=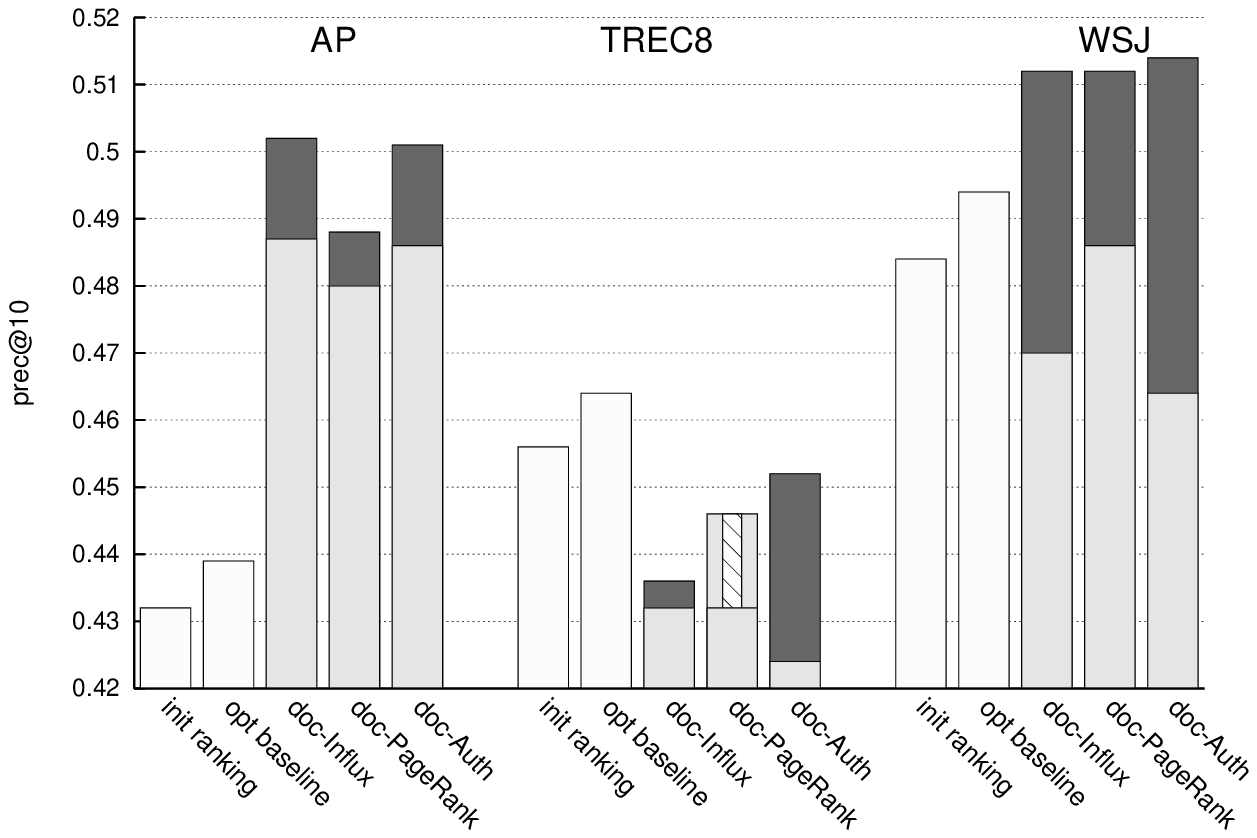,width=\graphwidth} \\
\psfig{figure=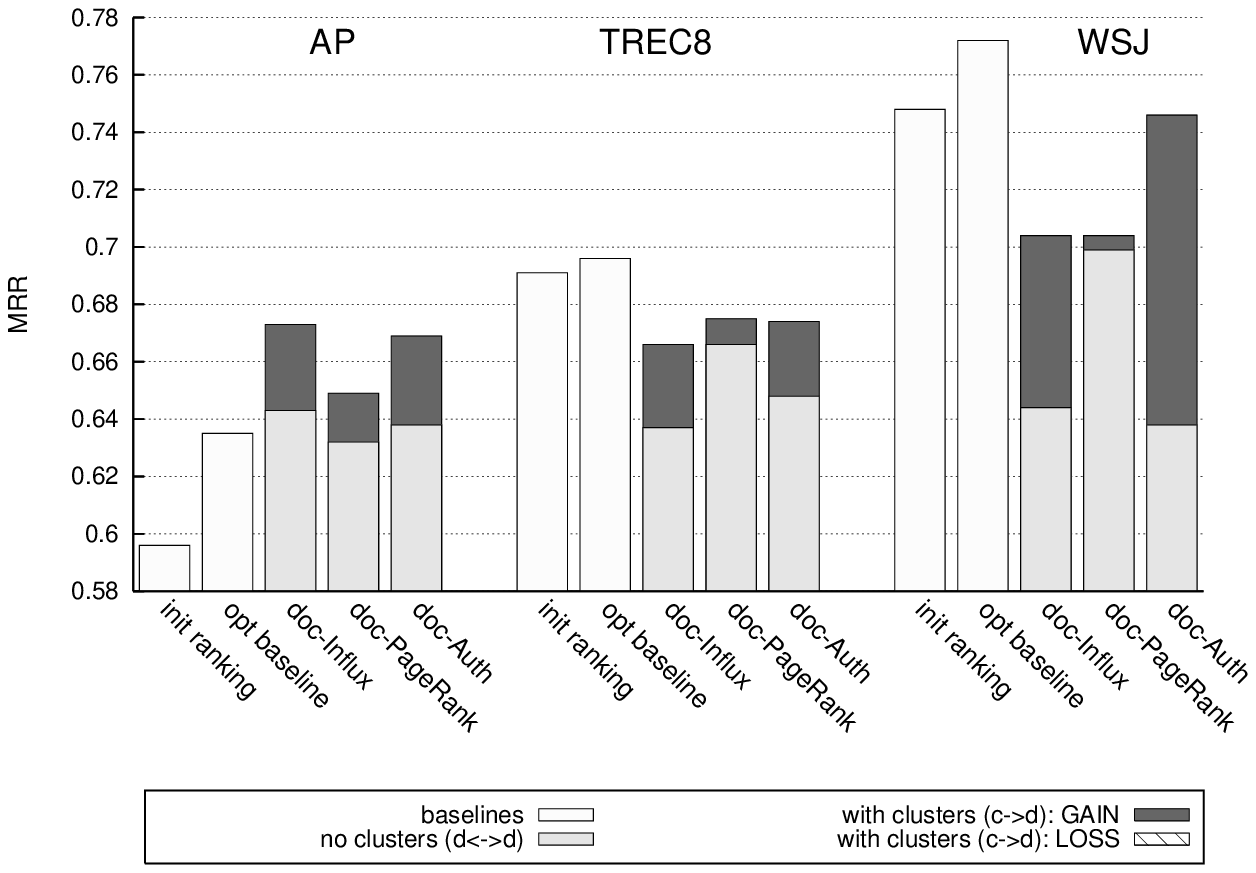,width=\graphwidth}
\end{tabular}
\caption{\label{fig:docAuth} All \reranking algorithms, as applied to either \dToDAbbrev graphs or
\cToDAbbrev graphs. }
\end{figure}
}

\paragraph*{Parameter selection for  graph-based methods}  There are
two motivations underlying our approach to choosing values for 
our algorithms' parameters
\cite{Kurland+Lee:05a}.  

First, we hope to show that
\rerankApproach can provide better results than the optimized
baselines even when initialized with a sub-optimal (yet reasonable)
ranking.  Hence, let the {\em initial ranking} be the document
ordering induced on the entire corpus by  $\inducedprob{{\ilmprob}^{\;[\dirichletParam_{1000}]}}{\doc}{\query}$, where 
$\dirichletParam_{1000}$ is the smoothing-parameter value optimizing
the average non-inter\-polated precision of the top $1000$
documents.  We set  $\topRetGroup$  to the top 50 documents in the initial ranking.

Second,
we wish to show that good results can be achieved without
a great deal of parameter tuning.  Therefore, we did not tune the
smoothing parameter for any of the language models used to determine graph
edge-weights, but rather simply set  $\dirichletParam=2000$ when
smoothing was required, following a prior suggestion \cite{Zhai+Lafferty:01a}.
Also, {\em the other free parameters' values were chosen so as to optimize
prec@5, regardless of the evaluation metric under
consideration.}\footnote{If two different parameter settings yield the
  same prec@5, we choose the setting {\em minimizing} prec@10 so as to
  provide a
conservative estimate of expected performance.  Similarly, if we have ties for both prec@5 and prec@10, we choose the setting
minimizing MRR.}  As a consequence, our prec@10 and
  MRR results are presumably not as high as possible; but the
  advantage of our policy is that we can see whether optimization with respect to a fixed
criterion yields good results no matter how ``goodness'' is measured.

Parameter values were selected from the following sets.
The graph ``out-degree''  $\numGenerators$:
$\{2,4,9,19,29,39,49\}$.  The cluster
size $\clustSize$:
$\{2,5,10,20,30\}$. 
The \prAlgorithm
damping factor $\dampFactor$:
$\{0.05, 0.1 \ldots 0.9, 0.95\}$.

\section{Experimental Results}
\label{sec:exp}

In what follows, when we say that
results or the difference between results are
``significant'', we
mean according to the two-sided Wilcoxon test at a confidence level of
$95\%$.

\subsection{\ReRanking by Document \PrestigeText}
\label{sec:docAuth}

\paragraph*{Main result} 
We first consider our main question: can we substantially boost the
effectiveness of \HITS by applying it to cluster-to-document graphs,
which we have argued are more suitable for it than the
document-to-document graphs we constructed in our previous work
\cite{Kurland+Lee:05a}?
The answer, as shown in Table \ref{tab:summaryTable}, is clearly
``yes'': 
we see that {\em moving to cluster-to-document graphs results in
substantial improvement for \HITS, and indeed boosts its results over
those for \prAlgorithm on document-to-document graphs.}

\paragraph*{Full suite of comparisons}
We now turn to Figure \ref{fig:docAuth}, 
which gives the
results for the \reranking algorithms \abbrevDocInfluxText,
\abbrevDocPRText and 
\abbrevDocAuthText as applied to either the
document-based graph \dToDAbbrev (as in \cite{Kurland+Lee:05a}) or the
cluster-document graph \cToDAbbrev. (Discussion of \abbrevDocHubText
is deferred to Section \ref{sec:furtherAnalysis}.)

To focus our discussion, it is useful to 
first
point out that 
in
almost all
of our nine evaluation
settings (3 corpora $\times$ 3 evaluation measures),
all 
three of
the \reranking algorithms perform better
when applied to
\cToDAbbrev graphs 
than to \dToDAbbrev graphs, as the number of
dark bars in Figure \ref{fig:docAuth}
indicates.
Since it is thus clearly useful to incorporate cluster-based
information, we 
will now mainly concentrate on
\cToDAbbrev-based algorithms.

The results for prec@5, the metric
for which the \reranking algorithms' parameters were optimized, show
that {\em all}
\cToDAbbrev-based algorithms outperform the prec@5-optimized baseline
--- 
significantly so for the AP corpus
--- even though
applied to a sub-optimally-ranked initial set.  (We hasten to point
out that while the initial ranking is  
always inferior to the corresponding 
optimized
baseline, the differences are
never 
significant.)  
In contrast, the use of \dToDAbbrev graphs never leads to
significantly superior prec@5 results.

We also observe in Figure \ref{fig:docAuth} that the
\abbrevDocAuthCtoD algorithm is always 
either the best of the
\cToDAbbrev-based algorithms
or clearly competitive with the best. Furthermore, 
pairwise comparison
of it to 
each 
of the \abbrevDocInfluxCtoD and \abbrevDocPRcToD algorithms
favors 
the \HITS-style \abbrevDocAuthCtoD algorithm
in a majority of the
evaluation settings.

We also experimented with a few  alternate graph-construction
methods, such as sum-normalizing the weights of edges out of nodes,
and found that the \abbrevDocAuthCtoD algorithm remained superior to
\abbrevDocInfluxCtoD and \abbrevDocPRcToD.  We omit these results due
to space constraints.

All in all,
these 
findings lead us to believe that
not only is it useful to incorporate information from clusters, but it
can be more effective to do so in a way reflecting the
mutually-reinforcing nature of clusters and documents, 
as the
\HITS algorithm does.

\subsection{\ReRanking by Cluster \PrestigeText}
\label{sec:clustAuth}

\newcommand{\clustAuthTableSize}{}
\newcommand{\clustAuthTableCaption}{\label{tab:clustAuthTable} 
Cluster-based \reranking.  Bold: best results per column.
Symbols {\it \statSymbolInit}, {\it \statSymbolOpt}, {\it \statSymbolCQL}:
results differ  signi\-ficantly from the initial ranking, 
optimized baseline, or (for the \reranking algorithms) \abbrevCQL
\protect \cite{Liu+Croft:04a}, respectively.
}
\begin{table*}[t]
\begin{flushleft}
\begin{tabular}{|l||c|c|c|c|c|c|c|c|c|}
\hline
& \multicolumn{3}{|c|}{AP}& \multicolumn{3}{|c|}{TREC8}& \multicolumn{3}{|c|}{WSJ}\\ \cline{2-10} 
 & {prec$@5$} & {prec$@10$} & {MRR} & {prec$@5$} & {prec$@10$} & {MRR} & {prec$@5$} & {prec$@10$} & {MRR} \\ \hline
\abbrevInit & $ .457 $ $ ^{}_{}$& $ .432 $ $^{}_{}$& $ .596 $ $^{}_{}$& $ .500 $ $ ^{}_{}$& $ .456 $ $^{}_{}$& $ .691 $ $^{}_{}$& $ .536 $ $ ^{}_{}$& $ .484 $ $^{}_{}$& $ .748 $ $^{}_{}$\\ \hline
opt. baselines & $ .465 $ $ ^{}_{}$& $ .439 $ $^{}_{}$& $ .635 $ $^{}_{}$& $ .512 $ $ ^{}_{}$& $ .464 $ $^{}_{}$& $ .696 $ $^{}_{}$& $ .560 $ $ ^{}_{}$& $ .494 $ $^{}_{}$& \mbox{\boldmath$ .772 $} $^{}_{}$\\ \hline
\abbrevCQL & $ .448 $ $ ^{}_{}$& $ .418 $ $^{}_{}$& $ .549 $ $^{\statSymbolInit\statSymbolOpt}_{}$& $ .500 $ $ ^{}_{}$& $ .432 $ $^{}_{}$& \mbox{\boldmath$ .723 $} $^{}_{}$& $ .504 $ $ ^{\statSymbolOpt}_{}$& $ .454 $ $^{\statSymbolInit\statSymbolOpt}_{}$& $ .680 $ $^{}_{}$\\ \hline
\hline
\hline
\abbrevClustInfluxDtoC & $ .511 $ $ ^{\statSymbolCQL}_{}$& \mbox{\boldmath$ .479 $} $^{\statSymbolCQL}_{}$& $ .619 $ $^{}_{}$& $ .524 $ $ ^{}_{}$& \mbox{\boldmath$ .478 $} $^{}_{}$& $ .681 $ $^{}_{}$& \mbox{\boldmath$ .568 $} $^{\statSymbolCQL}_{}$& \mbox{\boldmath$ .512 $} $^{\statSymbolCQL}_{}$& $ .760 $ $^{}_{}$\\ \hline
\abbrevClustPRdToC & $ .493 $ $ ^{}_{}$& $ .475 $ $^{\statSymbolCQL}_{}$& $ .595 $ $^{}_{}$& $ .496 $ $ ^{}_{}$& $ .444 $ $^{}_{}$& $ .683 $ $^{}_{}$& $ .528 $ $ ^{}_{}$& $ .490 $ $^{\statSymbolCQL}_{}$& $ .736 $ $^{}_{}$\\ \hline
\abbrevClustAuthDtoC & \mbox{\boldmath$ .533 $} $^{\statSymbolInit\statSymbolOpt\statSymbolCQL}_{}$& $ .478 $ $^{\statSymbolCQL}_{}$& \mbox{\boldmath$ .651 $} $^{\statSymbolCQL}_{}$& \mbox{\boldmath$ .532 $} $^{}_{}$& $ .460 $ $^{}_{}$& $ .714 $ $^{}_{}$& $ .552 $ $ ^{}_{}$& $ .478 $ $^{}_{}$& $ .757 $ $^{}_{}$\\ \hline
\end{tabular}

\caption{\clustAuthTableCaption}
\end{flushleft}
\end{table*}

We now consider the alternative, mentioned in Section
\ref{sec:algNames}, of using the \prestigeText scores for {\em
clusters} as an {\em indirect} means of ranking documents, in the
sense of identifying clusters that 
contain 
a high 
percentage of relevant
documents. 
Note
that the problem of automatically identifying 
such clusters 
in the \reranking setting has been
acknowledged 
to be a hard task for some time
\cite{Willett:85a}.
Nevertheless, as stated 
in Section \ref{sec:algNames}, we 
experimented with Liu and Croft's
general clusters-for-selection approach \cite{Liu+Croft:04a}: rank the
clusters, then rank the documents within 
each cluster by
$\dirichletLMTerm{\doc}{\query}{}$.  Our baseline algorithm,
\abbrevCQL, adopts 
Liu and Croft's specific proposal of the {\em CQL} algorithm
--- except that we employ overlapping rather than hard clusters ---
wherein clusters are ranked by the {\em query likelihood}
$\dirichletLMTerm{\clust}{\query}{}$ instead of one of our
\prestigeText scores.

Table \ref{tab:clustAuthTable} 
(which may appear on the next page)
presents the performance results. 
 Our
first observation is that the
\abbrevClustInfluxDtoC and \abbrevClustAuthDtoC algorithms are superior in a
majority of the relevant comparisons 
to the initial
ranking, the optimized 
baselines, and the \abbrevCQL algorithm,
where the performance differences with the latter sometimes achieve significance.

However, the performance
of the document-\prestigeText-based algorithm
\abbrevDocAuthCtoD
is better 
in a majority of the evaluation
settings
than that of any of the 
cluster-\prestigeText-based algorithms.
On the other hand,
it is possible that
the latter methods could be improved by a better technique for
within-cluster ranking.

To compare the effectiveness of \abbrevClustInfluxDtoC and
\abbrevClustAuthDtoC to that of \abbrevCQL in detecting clusters 
with a high percentage of relevant documents --- thereby neutralizing
within-cluster ranking effects --- we present in Table
\ref{tab:relClustAuthTableDensityCaption} the percent of 
documents in
the highest ranked cluster that are relevant. (Cluster size
($\clustSize$) was fixed to either 5 or 10 and out-degree
($\numGenerators$) was chosen to optimize 
the above percentage.)
Indeed, these
results clearly show that our best cluster-based algorithms are much
better than \abbrevCQL in detecting clusters 
containing
a high 
percentage of
relevant documents, 
in most 
cases to a significant degree.

\newcommand{\relClustAuthDensityTableSize}{\small}
\newcommand{\relClustAuthDensityTableCaption}{\label{tab:relClustAuthTableDensityCaption}
Average relevant-document percentage within the top-ranked cluster.
$\clustSize$: cluster size.
Bold: best results per column.
$\statSymbolCQL$: result  differs significantly from that of
\abbrevCQL, used in \protect \cite{Liu+Croft:04a}.
}
\begin{table}[h]
\hspace{-0.15in}
\relClustAuthDensityTableSize
\begin{tabular}{|l||l|l|l|l|l|l|}
\hline Cluster 
& \multicolumn{2}{|c|}{AP}& \multicolumn{2}{|c|}{TREC8}& \multicolumn{2}{|c|}{WSJ}\\ \cline{2-7} ranking 
 & \multicolumn{1}{c|}{$\clustSize$=$5$} & \multicolumn{1}{c|}{$\clustSize$=$10$} & \multicolumn{1}{c|}{$\clustSize$=$5$} & \multicolumn{1}{c|}{$\clustSize$=$10$} & \multicolumn{1}{c|}{$\clustSize$=$5$} & \multicolumn{1}{c|}{$\clustSize$=$10$} \\ \hline \hline
\veryAbbrevCQL & $ 39.2 $ $ ^{}_{}$& $ 38.8 $ $^{}_{}$& $ 39.6 $ $ ^{}_{}$& $ 40.6 $ $^{}_{}$& $ 44.0 $ $ ^{}_{}$& $ 37.0 $ $^{}_{}$\\ \hline
\veryAbbrevClustInfluxDtoC & $ 48.7 $ $ ^{\statSymbolCQL}_{}$& \mbox{\boldmath$ 47.6 $} $^{\statSymbolCQL}_{}$& $ 48.0 $ $ ^{}_{}$& $ 43.8 $ $^{}_{}$& $ 51.2 $ $ ^{\statSymbolCQL}_{}$& $ 48.0 $ $^{\statSymbolCQL}_{}$\\ \hline
\veryAbbrevClustAuthDtoC & \mbox{\boldmath$ 49.5 $} $^{\statSymbolCQL}_{}$& $ 47.2 $ $^{\statSymbolCQL}_{}$& \mbox{\boldmath$ 50.8 $} $^{\statSymbolCQL}_{}$& \mbox{\boldmath$ 46.6 $} $^{}_{}$& \mbox{\boldmath$ 53.6 $} $^{\statSymbolCQL}_{}$& \mbox{\boldmath$ 49.0 $} $^{\statSymbolCQL}_{}$\\ \hline
\end{tabular}

\caption{\relClustAuthDensityTableCaption}
\end{table}

\subsection{Further Analysis}
\label{sec:furtherAnalysis}

\paragraph*{Authorities versus hubs}
So far, we have only considered utilizing the authority scores that
the  \HITS algorithm produces.  
The chart below shows the effect of ranking entities by {hub} scores instead.  Specifically, 
the ``\docBasedRetrieval'' column compares
\mbox{\abbrevDocAuthCtoD} (i.e., ranking documents by authoritativeness) to
\mbox{\abbrevDocHubDtoC} (i.e., ranking documents by hubness);  similarly, the
``\clustBasedRetrieval'' column compares 
\mbox{\abbrevClustAuthDtoC} 
to \mbox{\abbrevClustHubCtoD}.
Each entry
depicts, in 
descending 
order of
performance (except for the one indicated tie) as one moves left to right,
those  \prestigeText scoring functions
that lead to an improvement over
the initial ranking:
$\authSymbol$ stands for ``authority'' and $\hubSymbol$ for ``hub''. Cases in which the improvement is 
significant are marked with a `\statSymbolCompTables'.
\newcommand{\compHubAuthTableSize}{}
\newcommand{\compHubAuthTableCaption}{\label{tab:compHubAuthTable}
Authority (\authSymbol) vs.  hub (\hubSymbol) scores as \prestigeText
measures. Each entry depicts, in 
non-ascending 
order of performance, the
\prestigeText scores that improve (`\statSymbolCompTables':
significantly so) over the initial ranking.  }
{\renewcommand{\caption}[1]{}
\begin{center}
\compHubAuthTableSize
\begin{tabular}{ll|*2{l}} \\ \hline
 \multicolumn{4}{c}{When do we improve the initial ranking} \\ 
  \multicolumn{4}{c}{by measuring the centrality of:} \\  
& & \multicolumn{1}{l}{\docBasedRetrieval}& \multicolumn{1}{l}{\clustBasedRetrieval}\\  \hline
& prec @5& \mbox{$\authSymbol^{*}$}$\hubSymbol$ & \mbox{$\authSymbol^{*}$}$\hubSymbol$ \\ {AP} 
& prec @10& \mbox{$\authSymbol^{*}$}$\hubSymbol$ & \mbox{$\authSymbol$}$\hubSymbol$ \\
& MRR& \mbox{$\authSymbol$}$\hubSymbol$ & \mbox{$\authSymbol$} \\ \hline& prec @5& \mbox{$\authSymbol$}$\hubSymbol$ & \mbox{$\authSymbol$}$\hubSymbol$ \\ {TREC8} 
& prec @10&  & \mbox{$\hubSymbol$}$\authSymbol$ \\
& MRR& \mbox{$\hubSymbol$} & \mbox{$\hubSymbol^{*}$}$\authSymbol$ \\ \hline& prec @5& \mbox{$\authSymbol$}$\hubSymbol$ & \mbox{$\authSymbol$}$\hubSymbol$ (tie)\\ {WSJ} 
& prec @10& \mbox{$\authSymbol$}$\hubSymbol$ & \mbox{$\hubSymbol$} \\
& MRR&  & \mbox{$\hubSymbol$}$\authSymbol$ \\ \hline\end{tabular}
\caption{\compHubAuthTableCaption}

\end{center}
} 
We see that  in many cases, hub-based \reranking does yield better performance than the initial
ranking.  But
authority-based \reranking
appears to be an even better choice overall.

\paragraph*{\HITS on \prAlgorithm-style graphs} Consider our comparison
of \abbrevDocAuthDtoD against \abbrevDocPRdTod.  As the notation
suggests, this corresponds to 
running \HITS and  \prAlgorithm on the same graph,
\dToDAbbrevText.  But an alternative interpretation \cite{Kurland+Lee:05a}
is that 
non-smoothed (or no-random-jump)  PageRank, as expressed by  Equation (\ref{eq:pagerank-concept}),
is applied to a {\em different} version of \dToDAbbrevText wherein the original
edge weights $\edgeWeight{\nodeFrom}{\nodeTo}$ have
been smoothed as follows:
\begin{equation}\edgeWeightSmooth{\nodeFrom}{\nodeTo}{\dampFactor} \definedas \frac{1 - \dampFactor}{|\arbGraphVertices|} + \dampFactor
\frac{\edgeWeight{\nodeFrom}{\nodeTo}}{\outwt{\nodeFrom}}
\label{eq:reweight}
\end{equation}
(we ignore nodes with no positive-weight out-edges to simplify
discussion, and omit the \dToDAbbrevText superscripts for clarity).  

How does \HITS perform on  document-to-document graphs that are
``truly equivalent'', in the sense of employing the above
edge-weighting regime, to those that \prAlgorithm is applied to?  One
reason this is an interesting question
is that \HITS assigns scores of zero to nodes that are not in the graph's
largest connected component (with respect to positive-weight edges,
considered to be bi-directional).  Notice that the original graph
may have several connected components, whereas utilizing $\smoothWtName{\dampFactor}$
 ensures that each node has a positive-weight directed edge to
every other node.  Additionally, the re-weighted version of \HITS has provable
stability properties \cite{Ng+Zheng+Jordan:01a}.

We found that in nearly all of our evaluation
settings for
document-to-document graphs  (three corpora $\times$ three evaluation metrics), \abbrevDocAuthDtoD
achieved better results using $\smoothWtName{\dampFactor}$ edge weights.  However, 
we cannot discount the possibility that the performance differences
might be due simply to the inclusion of the extra interpolation-parameter $\dampFactor$.  Moreover, in all but one case, the
improved results were still below those for \abbrevDocPRdTod (and
always lagged behind those of \abbrevDocAuthCtoD).

Interestingly, the situation is qualitatively different if we consider
\cToDAbbrev graphs instead. In brief, we applied a 
smoothing
scheme
analogous to that described above, but only to edges leading from a
left-hand node (cluster) to a right-hand node (document)\footnote{In
the \rightbip case, the ``$|\arbGraphVertices|$'' in Equation
(\ref{eq:reweight}) must be changed to the number of right-hand
nodes.}; we thus preserved the \rightbip structure.  Only in two of
the nine evaluation settings did this change cause an increase in
performance of \abbrevDocAuthCtoD over the results attained under the original
edge-weighting scheme, despite the fact that the re-weighting involves
an extra free parameter.  Thus, while we have already demonstrated in
previous sections of this paper that information about
document-cluster similarity relationships is very valuable, the
results just mentioned suggest that such information is more useful
in ``raw'' form.

\paragraph*{Re-anchoring to the query}

In previous work,
we showed that \prAlgorithm \prestigeText
scores induced over document-based graphs 
can
be used as a multiplicative
weight on document query-likelihood 
terms, the intent being to cope
with cases in which \prestigeText  in $\topRetGroup$ and relevance are not strongly
correlated
\cite{Kurland+Lee:05a}.  
Indeed, employing this technique on the AP, TREC8, and WSJ corpora, prec@5 increases from
$.519$, $.524$ and $.536$, to $.531$, $.56$ and $.572$
respectively.

The same modification could be applied to the \cToDAbbrev-based algorithms,
although it is not particularly well-motivated in the \HITS
case. While \prAlgorithm scores correspond to a stationary
distribution that could be loosely interpreted as a prior
\cite{Kurland+Lee:05a}, in which case multiplicative combination with
query likelihood is sensible, it is not usual to assign a
probabilistic interpretation to hub or authority scores.

Nonetheless, for the sake of comparison completeness, we applied this
idea to the \abbrevDocAuthCtoD algorithm, yielding the
following performance changes: 
from $.541$, $.544$, 
and
$.564$ to $.537$, $.572$ and $.572$ respectively.
These results are still as good as --- and for two corpora
better than --- those 
for \prAlgorithm as a multiplicative weight on query likelihood.
Thus, it may be the case that \prestigeText scores induced over a document-based graph are
 more effective as a multiplicative bias on 
query-likelihood
than as direct representations of  relevance in $\topRetGroup$ (see
also \cite{Kurland+Lee:05a});  but, modulo the caveat above, it seems
 that when \prestigeText is induced over 
cluster-based
\bipText graphs,
the correlation with relevance is 
much stronger, 
and hence 
this kind of \prestigeText
serves 
as a better ``bias'' on query-likelihood.

\section{Conclusion}
\label{sec:conc}
We have shown that leveraging the mutually reinforcing relationship between
clusters and documents to determine \prestigeText
is very beneficial not only for directly finding
relevant documents in an initially retrieved list, but also for
finding clusters of documents from this list that contain a high
number of relevant documents.

Specifically, we demonstrated the superiority of cluster-document
bipartite graphs to document-only graphs 
as the input to \prestigeText-induction algorithms.
 Our method for
finding ``authoritative'' documents 
(or clusters) using HITS over these
bipartite graphs results in state-of-the-art performance for document (and cluster)
\reranking.

\medskip
\scriptsize

\noindent {\bf Acknowledgments}~ We thank Eric Breck, Claire Cardie,
Oren Etzioni,
Jon Kleinberg, 
Art Munson, Filip Radlinski, Ves Stoyanov,
Justin Wick and the anonymous reviewers for valuable comments.  This
paper is based upon work supported in part by the National Science
Foundation under grant no. IIS-0329064 and an Alfred P. Sloan Research
Fellowship. Any opinions, findings, and conclusions or recommendations
expressed are those of the authors and do not necessarily reflect the
views or official policies, either expressed or implied, of any
sponsoring institutions, the U.S. government, or any other entity.

\end{document}